\documentclass[10pt,journal]{IEEEtranTCOM}

\usepackage[english]{babel}
\usepackage[usenames]{color}
\usepackage[cp1250]{inputenc}
\usepackage{amsfonts}
\usepackage{amsthm}
\usepackage{graphicx}
\usepackage{epsfig}
\usepackage{mathrsfs}
\usepackage{amsmath}
\usepackage{algorithm}
\usepackage{algorithmic}
\usepackage{moresize}
\usepackage{hyperref}

\theoremstyle{plain}

\begin{document}
\newcommand{\bea}{\begin{eqnarray}}
\newcommand{\eea}{\end{eqnarray}}
\newcommand{\be}{\begin{equation}}
\newcommand{\ee}{\end{equation}}
\newcommand{\beas}{\begin{eqnarray*}}
\newcommand{\eeas}{\end{eqnarray*}}
\newcommand{\bs}{\backslash}
\newcommand{\bc}{\begin{center}}
\newcommand{\ec}{\end{center}}
\def\SC {\mathscr{C}}

\title{Nearly accurate solutions for Ising-like models\\
using Maximal Entropy Random Walk}
\author{\IEEEauthorblockN{Jarek Duda}\\
\IEEEauthorblockA{Jagiellonian University,
Golebia 24, 31-007 Krakow, Poland,
Email: \emph{dudajar@gmail.com}}}
\maketitle

\begin{abstract}
While one-dimensional Markov processes are well understood, going to higher dimensions there are only a few analytically solved Ising-like models, in practice requiring to use relatively costly, uncontrollable and inaccurate Monte-Carlo methods. There is discussed analytical approach for e.g. $width\times \infty$ approximation of lattice, also exploiting Hammersley-Clifford theorem to generate random Gibbs/Markov field through scanning line-by-line using local statistical model as in lossless image compression. While its conditional distributions could be found with Monte-Carlo methods, there is discussed use of  Maximal Entropy Random Walk (MERW) to calculate them from approximation of lattice as infinite in one direction and finite in the remaining. Specifically, in the finite directions there is built alphabet of all patterns, then transition matrix containing energy for all pairs of such patterns is built, from its dominant eigenvector getting probability distribution of pairs of patterns in Boltzmann distribution of their infinite sequences, which can be translated into local statistical model for line-by-line scan. Such inexpensive models, requiring seconds on a laptop for attached implementation and directly providing probability distributions of patterns, were tested for mean entropy and energy per node, getting maximal $\approx 0.02$ error from analytical solution near critical point, which quickly improves to extremely accurate e.g. $\approx 10^{-10}$ error for $J\approx 0.2$.
\end{abstract}
\section{Introduction}
\begin{figure}[t!]
    \centering
        \includegraphics[width=8.5cm]{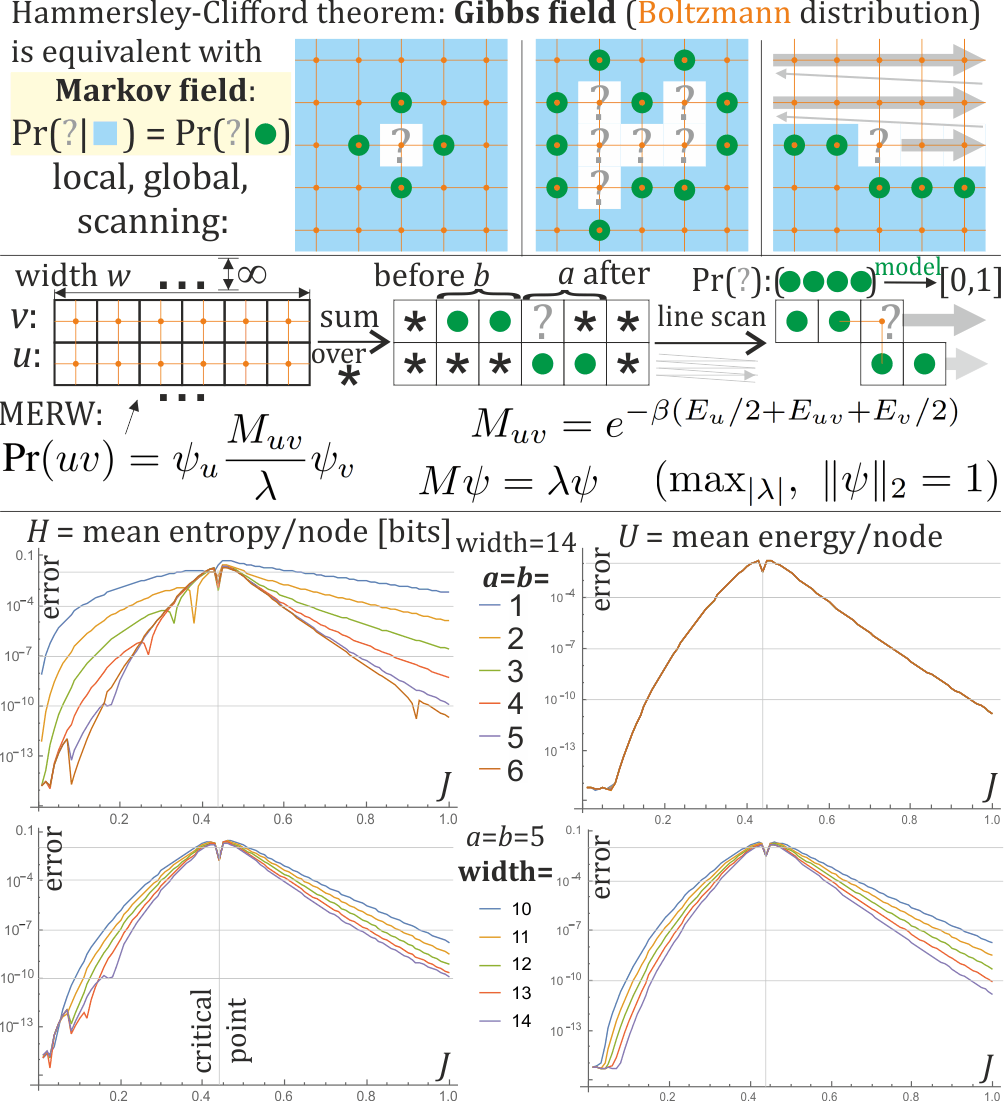}
        \caption{\textbf{Top}: we focus on Gibbs fields - defined by Boltzmann probability distribution $\textrm{Pr}(s) \propto e^{-\beta E(s)}$ of 2D lattice configurations $s:\mathbb{Z}^2\to \mathcal{A}$ with short-range interactions: energy $E$ being sum of functions of values in each node and edge (visualized with orange color). By Hammersley-Clifford theorem, Gibbs field is equivalent with Markov field: in which conditional probability distributions based on values in some region (blue), depend only on values in the boundary of this region (green). \textbf{Below}: we use the presented MERW formula to find probability distribution of $(u,v)$ pairs of neighboring patterns of finite $w$ width - getting analytical solution for Boltzmann distribution among their infinite sequences in vertical direction. Summing over the '*' positions, and dividing by sum over '?' values, we get the model: probability of '?' value based on a few neighboring values marked green. We can use such calculated local conditional probability distribution model e.g. to generate random field through such scanning line-by-line, calculate parameters like mean entropy or energy per node, probability distribution of patterns, data compress such configuration or store information in it. \textbf{Bottom}: errors (differences from accurate values) of calculated entropy (left) and energy (right) per node found this way for various $J=J_v=J_h$ coupling parameters ($\beta=1, \mu=0$), using various model size (up) and widths used to calculate the model (down). These errors reach $\approx 0.02$ near critical point $J\approx 0.44$, but reduce orders of magnitude if moving away from this point. We can see characteristic beak near critical point, which might be useful to localize it in general case.}
        \label{ising}
\end{figure}
Gibbs fields are the basic models of statistical mechanics, condensed matter physics. In 1971 through Hammersley-Clifford theorem~\cite{HC}, they turned out equivalent with providing different perspective Markov fields, stimulating both fundamental information theory research as being interesting topic very far from complete understanding, but also having potential for new practical tools and applications especially around image analysis, segmentation, compression~\cite{markov1, markov2}, or maybe for storing information in constrained media like Bit Patterned Media Recording~\cite{BPMR}.

As there are only few models with known analytical solution like zero-field 2D Ising model~\cite{onsager} awarded Nobel prize, in practice there are mainly used Monte-Carlo methods~\cite{MC} which are costly to provide uncorrelated fields, agreement with desired probability distribution, high accuracy for estimated properties, can get trapped in local minima preventing from exploring full configuration space~\cite{opt}.

Hence it is valuable to develop intermediate methods e.g. analytical for approximated problem, like discussed here visualized in Fig. \ref{ising}, which combines two concepts that can be also applied separately:
\begin{itemize}
  \item Markov fields allow to simplify calculation of conditional probability distribution to be based only on local situation, what is very convenient from perspective of generating them by scanning line-by-line as in lossless image compression like LOCO-I~\cite{loco}.
  \item Such model of local conditional distributions requires to know probability distribution of patterns - they could be found with Monte-Carlo, but it is costly to get high accuracy this way. Here they are found with Maximal Entropy Random Walk (MERW) providing analytical solution for approximation of lattice to infinite in one direction and finite in the remaining.
\end{itemize}
Presented methodology was originally introduced~\cite{org} by the author in 2006 for 2D Fibonacci coding (Hard Square) problem - for storing information in lattice of bits where it is forbidden to use two neighboring '1's, what could e.g. allow to improve HDD capacity. Statistical model found with MERW allowed to work $\approx 10^{-9}$ bits/node (for $a=b=5$ here) below theoretical $\approx 0.5878911617753406$ bits/node~\cite{baxter} entropy threshold, and first Asymmetric Numeral Systems were introduced there for such (reversed) entropy coding purpose. This article expands it from uniform distribution among allowed patterns into Boltzmann distribution for general Ising-like models. Appendix A contains example implementation, Appendix B contains divagations on using realisation of Ising-like model for computing purposes.

\section{Problem and methodology}
\subsection{Gibbs fields with Ising model as practical example}
Information theoretic discussion of the proposed methodology can be found in~\cite{org} from perspective of 2D Fibonacci/Hard Square problem  forbidding some patterns. Here we would like to expand it to more general Boltzmann distribution of patterns in Gibbs field.

We focus on (infinite) $d$-dimensional lattice $X=\mathbb{Z}^d$, being interested in its space of configurations: $s:X\to \mathcal{A}$ for some alphabet $\mathcal{A}$, e.g. $\{0,1\}$ or $\{-1,1\}$. We focus on discrete alphabets, but there are also considered continuous.

We would like to consider Boltzmann distribution among such configurations:
\be \textrm{Pr}(s)\propto \exp(-\beta E(s)) \ee
where $\beta$ can be fixed here to $\beta=1$, $E(s)$ is called energy of configuration and usually is defined in a translationally invariant way, for example in popular 2D Ising model we will focus on here: $\mathcal{A}=\{-1,1\}$, energy $E(s)=$
\be -\mu \sum_{x,y\in\mathbb{Z}} s_{x,y} - J_h\sum_{x,y\in\mathbb{Z}} s_{x,y} s_{x+1,y} - J_v\sum_{x,y\in\mathbb{Z}} s_{x,y} s_{x,y+1}\label{is}\ee
where $\mu$ corresponds to external field, $J_h, J_v$ are horizontal and vertical coupling constants, usually the same: the presented tests are for $J_h=J_v=J$, $\beta=1$, $\mu=0$.

Some properties of interest to be found are average entropy, energy, value per node - there are known analytical formulas for $\mu=0$ case~(\cite{onsager, baxter}) used in tests here. As the lattice is infinite, such local averages can be defined by limit of averages from finite size lattices. Here it is avoided by imagining that all values were chosen by translationally invariant conditional probability distribution model: for line-by-line scanning, what allows to obtain averages from probability distribution of its local situations.

From information theory perspective it is also valuable to be able to use the presented approach to calculate probability distributions of patterns for such ensembles, what defines their Markov type~\cite{types}.

All such questions can be approximately answered and in practice usually are using Monte-Carlo methods like Metropolis-Hastings~\cite{MC}. They generate random fields on finite lattice e.g. with cyclic boundary conditions by repeating large number of times: take random position and randomly modify its value or not accordingly to calculated probability depending on its neighbors. While we have certainty of asymptotically getting field from the assumed probability distribution, in practice only finite numbers of steps are used - bringing difficult questions of probability distribution it has actually used and autocorrelations with the previously generated field, in addition to contribution of its approximation with finite lattice.

Additionally, even if being from perfect desired distribution, estimation of probability distribution of patterns from random samples has error decreasing with square root of the sample size - making it impractical to get e.g. $\sim 10^{-10}$ accuracy, which often can be inexpensively achieved with the discussed approach as in evaluation in Fig. \ref{ising}.
\subsection{Hammersley-Clifford theorem and Markov property}
While Gibbs fields seem not very convenient to understand local statistical behavior, fortunately through Hammersley-Clifford theorem~\cite{HC} they turn out equivalent with Markov fields, which generalize property from Markov processes into multidimensional case, can be written e.g. as:
\be \textrm{Pr}(s_{x,y}|X\backslash (x,y))=\textrm{Pr}(s_{x,y}|\mathcal{N}(x,y)) \ee
where $\mathcal{N}$ is neighborhood accordingly to the used interaction/constraints in energy formula, e.g. 4 neighbors in (\ref{is}) of Ising model

This is local Markov property, visualized in the left diagram in top of Fig. \ref{ising}. Applying it multiple times we get global formulation from the middle diagram: conditional distribution in some set of nodes depends only of values at its boundary according to the used interaction/constriants.

Finally we can also use it for scanning (right diagram): generate random field through taking succeeding line-by-line random values like in lossless image compression. Markov property says that conditional distribution based on already fixed values in fact depends only on the boundary values. In $\mathbb{Z}^2$ lattice this boundary is infinite, but we can approximate using local: a few values before and after - correspondingly $b$ and $a$ values as in this diagram.

For such fixed $b$ values before and $a$ after, we can put the entire behavior into size $|\mathcal{A}|^{a+b}$ table - of '?' conditional probability based on all $a+b$ values. To find such model we need to estimate probability distribution of all size $a+b+1$ patterns including '?', then divide it by sum over all values of '?'. In Fig. \ref{ising} error plot we can see that while calculated entropy strongly depends on such $a,b$ context size, energy nearly does not.

We could find such pattern distributions with Monte-Carlo, but it would be costly to get decent accuracies this way - here we use MERW instead.

\subsection{Maximal Entropy Random Walk (MERW)}
MERW~(\cite{org,MERW1}) analytically solves general 1D problem, is further applied to graph of all width $w$ patterns. Originally for a graph adjacency matrix $M$ it provides stochastic matrix assuming uniform distribution of paths on this graph, or equivalently maximizing entropy rate. As sketched below, it can be generalized from uniform to Boltzmann distribution~\cite{MERW2} by replacing adjacency matrix with:
\be M_{uv}=\exp(-\beta(E_u/2+E_{uv}+E_v/2)) \label{trans} \ee
where $E_u$ is energy of vertex/pattern $u$, $E_{uv}$ of $u-v$ edge/interaction. In statistical physics it is referred as transition matrix~\cite{baxter} and averaged properties are calculated from its eigenvalues. Here we would like to find stochastic model, what requires its dominant eigenvector instead:
\be M\psi=\lambda \psi \qquad \textrm{for maximal }|\lambda|,\quad \sum_i \psi_i^2=1 \label{eig}\ee
From Perron-Frobenius theorem this eigenvector should be unique and can be chosen as real, allowing to approximate $M^p\approx \lambda^p \psi \psi^T$ for large powers $p$. From the other side from (\ref{trans}), $M^{2p}$ can be seen as sum (partition function) over all length $2p$ paths using Boltzmann distribution. Fixing their central position to $u$ and performing $p\to \infty$ limit, we get probability distribution of patterns $\textrm{Pr}(u)\propto (\psi_u)^2$. Analogously fixing its neighboring two central positions to $u,v$ and performing $p\to \infty$ limit we get $\textrm{Pr}(u,v)\propto \psi_u M_{uv} \psi_v$. Normalizing them to sum to 1, thanks to $\sum_u (\psi_u)^2=1$ and $M\psi=\lambda \psi$, we get probability distributions for one and two neighboring vertices/patterns:
\be \textrm{Pr}(u)=(\psi_u)^2\qquad\qquad \textrm{Pr}(u,v)= \psi_u \frac{M_{uv}}{\lambda} \psi_v \label{merw} \ee

\subsection{Calculating the model}
\begin{figure}[t!]
    \centering
        \includegraphics[width=8.5cm]{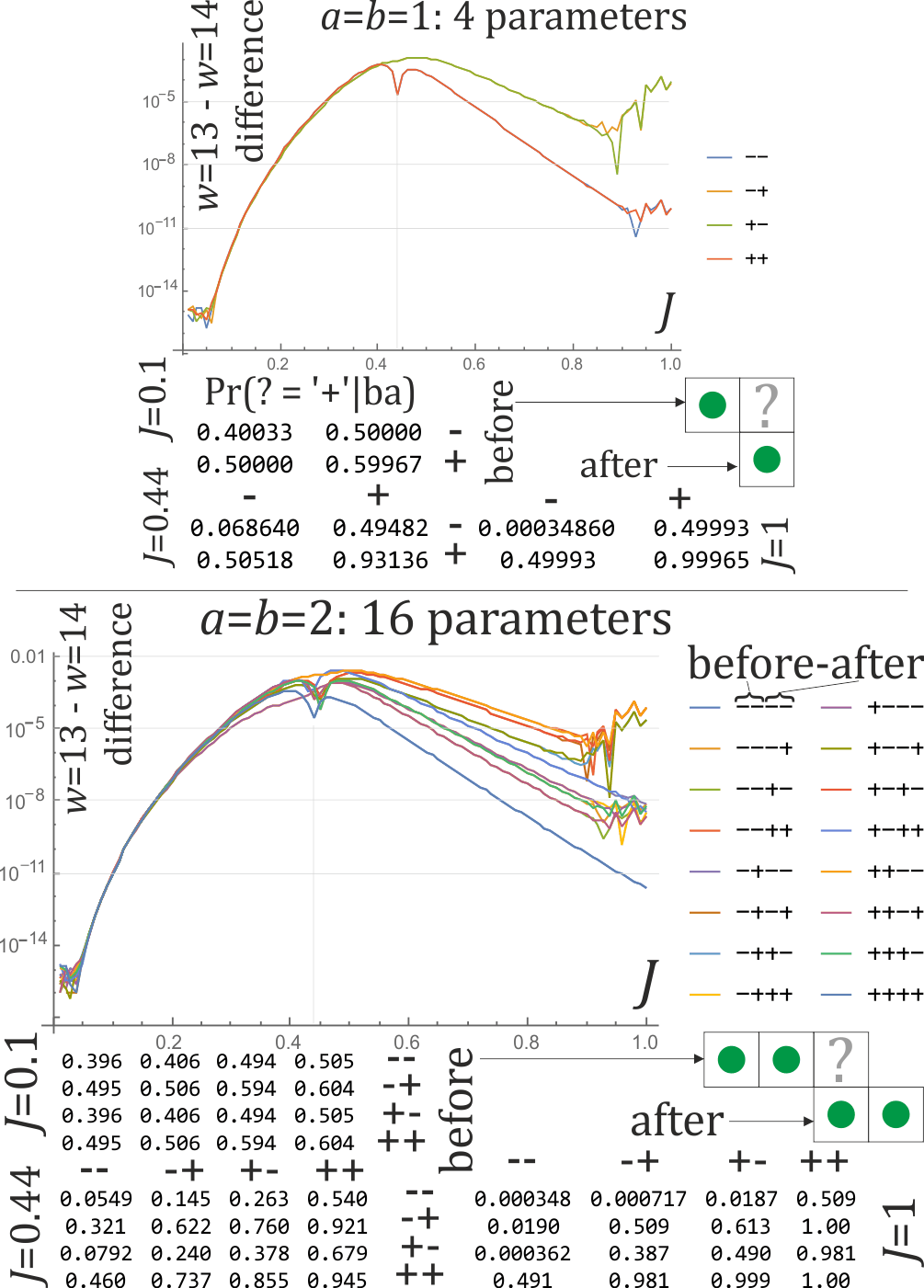}
        \caption{Plots: differences between calculated  parameters of models ($\textrm{Pr}(?=+|\textrm{before, after})$) using width=14 and width= 13, for $a=b=1$ model with $2^2=4$ parameters (up) and $a=b=2$ model with $2^4=16$ parameters (down) to evaluate their accuracies. We can see that especially above the critical point ($J\approx 0.44$) parameters for opposite neighboring spins have lower accuracies. Characteristic beak near critical point suggests that they could be localized this way. Below plots there are parameters (probabilities of using '+') for width 14 and $J=0.1, 0.44, 1$ cases. }
        \label{models}
\end{figure}
Formula (\ref{merw}) allows us to calculate probability distribution of pairs of neighboring values assuming Boltzmann distribution among their infinite sequences.

To apply it to a multidimensional lattice case, we can approximate the remaining directions with finite width $w$ like in Fig. \ref{ising}, getting random walk in large size $|\mathcal{A}|^w$ set of states $\mathcal{A}^w$. We can assume cyclic boundary conditions - approximating $\mathbb{Z}^2$ lattice with infinite length cylinder surface. Otherwise we approximate with $w\times \infty$ rectangle.

Now for the assumed interaction model we need to calculate individual energy of each $\mathcal{A}^w$ pattern, and interaction energy between all their pairs, constructing transition matrix (\ref{trans}). Calculating its dominant eigenvector we can find probability distribution of pairs $\textrm{Pr}(u,v)$ using (\ref{merw}).

Now we can calculate the model by summing over values in unused positions - marked by stars in Fig. \ref{ising}, and dividing by sums over the searched position '?'. Appendix contains optimized implementation of this procedure in Wolfram Mathematica, Fig. \ref{models} contains examples of such models.

\subsection{Applying the scanning model}
Discussed model provides conditional probability distribution for the currently considered position, based on local already chosen values in scanning line-by-line.

For low dimension and short range interactions it can be inexpensively calculated with discussed MERW-based approach, providing nearly accurate values unless being close to a critical point. In remaining  cases we can always search for such model with Monte-Carlo, however, getting high accuracy might become computationally very costly.

Having such model, a basic application is calculating average energy, entropy, value per node, e.g. by averaging such properties for currently chosen value over estimated probability distribution of considered size $a+b$ context - e.g. as realized in Appendix.

Another application is generation of random field using such model through line-by-line scan, for boundary values using model with reduced context. As being generated from scratch, we get practically uncorrelated configuration this way. There remains question of its agreement with assumed distribution, in any case there can be later applied some reduced number of Monte-Carlo steps to improve this distribution - details of savings which can be obtained this way need further investigation.

Other possible applications of such scanning models is data compression of random fields, or storing information in such constrained media by using entropy coder instead of taking random values, as in the original motivation for the discussed approach~\cite{org}.

\subsection{Alternative MERW calculation of model}
MERW formulas (\ref{merw}) for probability distribution of symbols/patterns and their pairs can be analogously extended to formulas for longer: length $l$ sequences of symbols/patterns by using $l-1$ appearances of $M/\lambda$ matrix with corresponding indexes:
\be \textrm{Pr}(u_1 u_2 \ldots u_l)=\psi_{u_1} \frac{M_{u_1 u_2}}{\lambda}\frac{M_{u_2 u_3}}{\lambda}\ldots  \frac{M_{u_{l-1} u_l}}{\lambda}\psi_{u_l} \label{seq1}\ee
In practice we might want to allow multiple patterns in above intermediate indexes, e.g. all with fixed '+1' in some node. This summation can be generally written by building $\Pi_i$ matrices with '1' on diagonal for all allowed cases for given position, and zeros beside:
\be \textrm{Pr}(\Pi_1 \ldots \Pi_l)= \psi^T\, \Pi_1\, \frac{M}{\lambda}\, \Pi_2\, \frac{M}{\lambda} \ldots \frac{M}{\lambda}\, \Pi_l\, \psi \label{seq}\ee
Considered various $\Pi_i$ for a given position should sum over all alternatives to identity matrix. The $\psi$ at the ends can be imagined as a result of infinite sequence of multiplication of $M/\lambda$, of propagators from both infinities.

We could use the above formula for alternative calculation of model parameters: by placing considered nodes (green dots and '?' in Fig. \ref{ising}) in vertical instead of horizontal way, and using (\ref{seq}) formula for $l=a+b-1$ length.

Intuitively it could lead to more accurate probabilities thanks to being further from boundaries, however, it would require many multiplication operations for very large dense matrices - what is computationally very expensive.

\section{Continuous case}
\begin{figure}[b!]
    \centering
        \includegraphics[width=8.5cm]{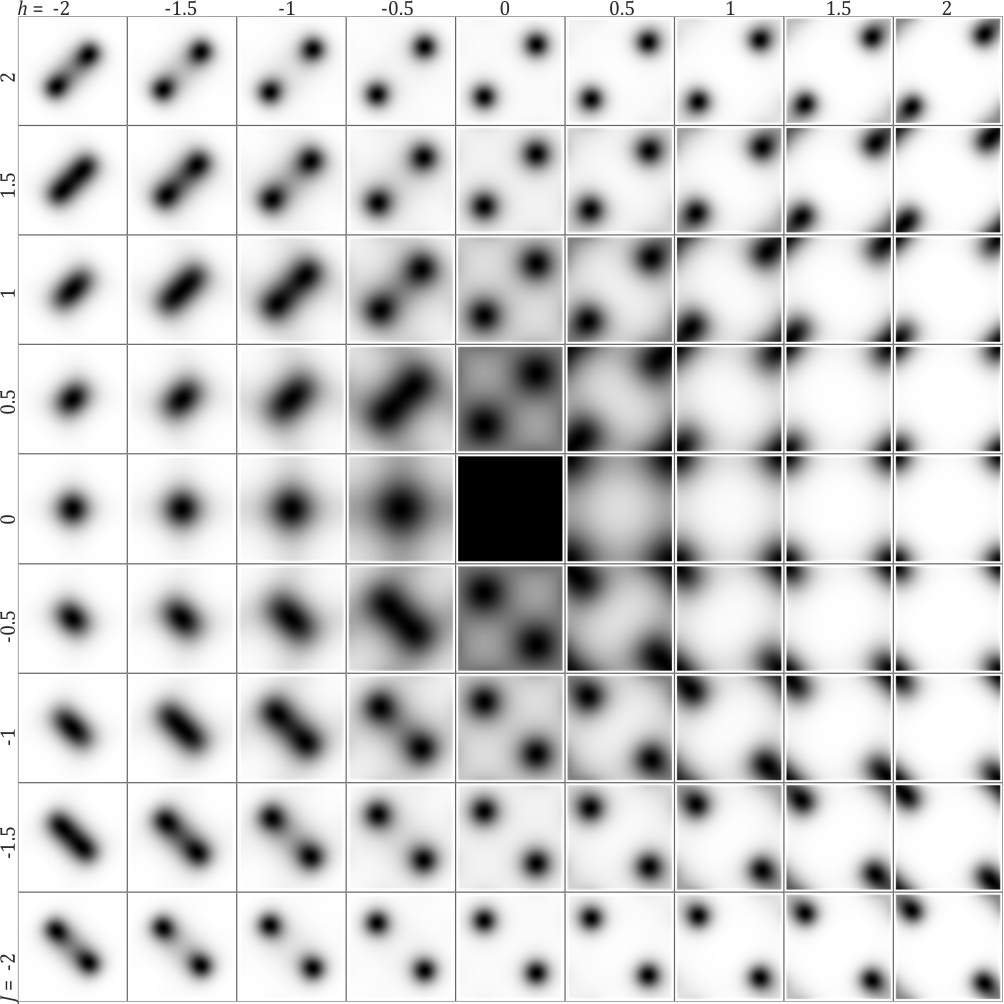}
        \caption{Joint distributions of $(\alpha_i, \alpha_{i+1})$ for neighboring spins in discussed 1D transverse-field Ising model (\ref{tising}) approximated with MERW for 100x100 lattice and various $J,h$ parameters. In the center there is no interaction, hence we can see uniform joint distribution. For $h=0$ we can see (anti)ferromagnetism with additional thermal wobbling around (anti)aligned spins. For $J=0$ we can see thermal wobbling around the energetically preferred spin directions. Generally we can observe intermediate situations from competition between these two tendencies.}
        \label{tis}
\end{figure}
We were working on discrete alphabet: spins up or down. However, in reality this direction of magnetic dipole of spin should be rather a continuous parameter, behaving accordingly to dipole-dipole interaction: energy for $\mathbf{m}_1, \mathbf{m}_2$ dipoles in distance $\mathbf r \neq 0$\footnote{\url{https://en.wikipedia.org/wiki/Magnetic_dipole\%E2\%80\%93dipole_interaction}}:
$$ H = -\frac{\mu_0}{4\pi|\mathbf r|^3}\left( 3 (\mathbf{m}_1\cdot\hat{\mathbf r})(\mathbf{m}_2\cdot \hat{\mathbf r})- \mathbf{m}_1\cdot \mathbf{m}_2\right) $$
Popular nontrivial case is transverse-field Ising model~\cite{ti}: 1D lattice of spins in $xz$ 2D plane, only the nearest neighbor interactions additionally restricted to single direction, with external field described by $h$:
\be E(s)=-J \sum_i s^z_i s^z_{i+1} - h \sum_i s^x_i  \label{tising}\ee
While it is usually analyzed in quantum way, let as look at it classically here: parameterize spin angle in $xz$ plane e.g. as
\be s^x_i =\cos(\alpha_i)\qquad s^z_i = \sin(\alpha_i)\ee
Representing spin sequence as sequence of spin angles $s\equiv (\alpha_i)_{i\in \mathbb{Z}}$, we can assume Boltzmann ensemble among such sequences $\textrm{Pr}(s)\propto \exp(-\beta E(s))$, for simplicity using $\beta=1$ here as we can put it into $J,h$ parameters.

We can analogously use MERW here e.g. for discrete lattice of $\alpha$, find transition matrix $M$ for such discretized angles, solve the dominant eigenequation, and use MERW formula e.g. for joint distribution of $(\alpha_i, \alpha_{i+1})$ pairs here, or conditional distribution for continuous Markov process of angles. Figure \ref{tis} shows such joint distributions calculated with below Mathematica implementation for \verb"lat"=100 lattice size (symmetrizing energy):
\begin{small}
\begin{verbatim}
en = -J*Sin[a]*Sin[b] - h*(Cos[a]+Cos[b])/2;
M = Table[Exp[-en], {a,2Pi/lat,2.Pi,2Pi/lat}
    ,{b,2Pi/lat,2.Pi,2Pi/lat}];
es = Eigensystem[M, 1];
pr = KroneckerProduct[es[[2, 1]],es[[2, 1]]]
     * M/es[[1, 1]];
\end{verbatim}
\end{small}

While in theory as  previously we could generalize it to more complex cases like 2D lattice, the exponential growth of matrix size would make it rather impractical - suggesting to search for approximations.

For example, in discussed lattice approximation each discretized angle corresponds to density of position being 1 around this $\alpha$ and 0 otherwise. Maybe we could use a different basis of functions here to represent density as a linear combination, e.g. of orthonormal polynomials, or Fourier-like sines-cosines to include periodicity of $\alpha$ here, or spherical harmonics for spin direction in 3D etc.

Let us consider using density as $\rho(x) = f_0 + \sum_{i>0} a_i f_i(x)$ for some $(f_i)$ orthonormal family, with $f_0$ guarding normalization to integrate to 1, e.g. $f_0 =1/2\pi$ for angles. Like in \cite{HCR}, we could estimate coefficients as $a_i = \textrm{mean over sample of }f_i(x)$ for random sample obtain e.g. with Monte-Carlo.

It would be valuable to get such coefficients directly from MERW-like model, e.g. for $E_{ij}=\int E(x,y) f_i(x) f_i(y) dx dy$ energy contribution for given neighboring $(i,j)$ pair in such basis, from $E(x,y)$ energy for the original variables. MERW for $M_{ij}=\exp(-\beta M_{ij})$ transition matrix would see it as ensemble of sequences of basis functions, finding optimal probability distribution of their patterns. It resembles quantum approach and is planned to be explored in the future.

\section{Discussion and further work}
There was discussed perspective for Gibbs/Markov fields as being generated through scanning line-by-line like in lossless image compression. Hammersley-Clifford theorem allows to optimize the necessary context for conditional probability distributions. MERW-based approach allows to inexpensively calculate quite accurate models at least in some situations. Having such model we can e.g. estimate parameters, probability distributions of patterns, inexpensively generate nearly uncorrelated random fields, data compress them or store information in such constrained media.

While this approach can be adapted to various short-range interactions in 2D (also long in pattern direction), it quickly becomes impractical for longer range interactions and higher dimensions due to exponentially growing space of possible patterns. It might be possible to overcome e.g. by working on some spaces of features or classes of abstraction of possible patterns. Even more difficult question regards continuous case, briefly discussed in the previous section, requiring to model continuous conditional probabilities this way, what could be handled e.g. by discretization or representing density in some optimized basis.
\bibliographystyle{IEEEtran}
\bibliography{cites}

\appendices
\section{Implementation for Ising model}
This Appendix contains optimized Wolfram Mathematica implementation of discussed method, also calculation of average entropy, energy and magnetization, and analytical formulas to find their exact values for tests. For width=13 it needs about 3 seconds, for width=14 about 15 seconds. For higher widths memory requirements quickly grows due to the needed dense $2^w\times 2^w$ size matrix \verb"prob" which is used for both transition matrix and pair probabilities here.
\begin{ssmall}
\begin{verbatim}
(* findprob - find Pr(u,v) of pairs of stripes *)
(* needs Jv, Jh, mu, beta, width, cyclic *)
findprob := ( patn = Power[2,width];
  pat = Table[2 IntegerDigits[i,2,width]-1.,{i, 0, patn-1}];
  tmp = Transpose[pat]; If[cyclic, AppendTo[tmp, tmp[[1]]]];
  prod = Transpose[tmp[[1 ;; -2]]*tmp[[2 ;; -1]]];
  paten =                          (*  energy of patterns *)
   Table[-Jh*Total[prod[[i]]]-mu*Total[pat[[i]]],{i, patn}];
          (* calculate transfer matrix as prob: *)
  prob = KroneckerProduct[Table[1., patn], paten/2];
  prob += Transpose[prob] - pat.Transpose[Jv*pat];
  prob = Exp[(-beta)*prob];      (* into transfer matrix *)
  {{lam},{psi}}=Abs[Eigensystem[prob,1,Method ->"Arnoldi"]];
  prob*=KroneckerProduct[psi/lam,psi];)  (* found Pr(u,v) *)

(* find model: pr. of '?' based on bef 'b's before  *)
(* [mid-bef,mid-1] and aft 'a's after [mid,mid+aft-1]: *)
(* ***bbb?***** - next pattern (j) *)
(* ******aaa*** - previous pattern (i) in (ij) pairs *)
h[p_]:=-p*Log[2,p]-(1-p)Log[2,1-p];    (* Shannon entropy *)
findmodel := (tpat = Round[Transpose[pat]]/2 + 1/2;
  mid = Ceiling[width/2 + 1];          (* position of '?' *)
  ipat=Table[Power[2,i],{i,bef,0,-1}].tpat[[mid-bef;;mid]]+1;
  jpat=Table[Power[2,i],{i,aft-1,0,-1}].tpat[[mid;;mid+aft-1]]+1;
  rprob =                             (* summing over '*' *)
     SparseArray[Table[{ipat[[i]],i}->1., {i, patn}]].prob.
     SparseArray[Table[{j, jpat[[j]]} -> 1., {j, patn}]];
  cprob = rprob[[1 ;; -1 ;; 2]] + rprob[[2 ;; -1 ;; 2]];
  model = rprob[[2 ;; -1 ;; 2]]/cprob;  (* find the model *)
  H = Total[Total[cprob*h[model]]];   (* entropy per node *)
  mag = Total[Total[                     (* magnetization *)
  (rprob[[2 ;; -1 ;; 2]] - rprob[[1 ;; -1 ;; 2]])*cprob]];
  U = Total[Total[Table[curs=2BitAnd[i,1]-1;    (* energy *)
       hs=BitAnd[i,2]-1; vs=2 BitShiftRight[j, aft- 1] - 1;
       -curs(mu + Jv*vs + Jh*hs)
       ,{i,0,Power[2,bef+1]-1},{j,0,Power[2,aft]-1}]*rprob]];
  {U, H, mag});

 (* caluclates accurate values for U and H *)
 accUH[J_, beta_: 1] := (If[J == 0, {0., 1.},
   k = 1/Power[Sinh[2*beta*J], 2];
   U = -J*Coth[2*beta* J]
    (1 + (2/Pi)*(2 Power[Tanh[2*beta*J], 2] - 1)
    NIntegrate[1/Sqrt[1-4k*Power[1+k,-2]*Power[Sin[th], 2]],
    {th, 0, Pi/2}]);
   F = -Log[2]/(2*beta) - (1/(2 Pi*beta))
    NIntegrate[Log[Power[Cosh[2*J*beta], 2] +
         Sqrt[1+Power[k,2]-2 k*Cos[2*th]]/k], {th, 0, Pi}];
   {U, beta (U - F)/Log[2]}])

(* example of application *)
Jv = Jh = 1; mu = 0; beta = 1;
width = 10; cyclic = True;
findprob;
bef = 3; aft = 3;
Print[accUH[Jv, beta], " accurate {U,H}, found:"];
findmodel (*return {U,H,magnetization}*)
\end{verbatim}
\end{ssmall}
\section{(Wick-rotated) Ising Quantum Computing (IQC)}
While quantum mechanics is equivalent with Feynman path integrals of  $\gamma(t)$ paths in time direction, discussed Ising-like models (e.g. for $\mathcal{A}^w$ width $w$ stripes of spins) are generally assumed to use Boltzmann ensemble among discrete sequences: $\gamma_x$ this time in space, being spatial realization of MERW. Feynman and Boltzmann path ensembles mathematically differ by Wick rotation, sharing many common features like analogous stationary probability distribution with localization property and Born rule, e.g. predicting $\rho(x)\propto \sin^2(\pi x)$ stationary density for dynamics in $[0,1]$ range, instead of $\rho=1$ for standard diffusion.

MERW has $\textrm{Pr}(u)=(\psi_u)^2$ Born rule-like behavior literally from symmetry, as one amplitude $\psi$ comes from $\lim_{p\to \infty} (M/\lambda)^p=\psi \psi^T$ propagator from left, second from right. Born rule's square is essentially different from axioms of standard (Kolmogorov) probability theory, hence Bell-like inequalities derived in the latter are not necessarily satisfied in the former. For example Mermin's inequality $\textrm{Pr}(A=B)+\textrm{Pr}(A=C)+\textrm{Pr}(B=C) \geq 1$ for $ABC$ binary variables, analogous to "tossing 3 coins, at least 2 give the same", can be violated by QM formalism.
It is discussed in \cite{fqm}, containing example of such MERW violation of this inequality thanks to Born rule based probabilistics, which might be realizable with discussed Ising-like systems.

\begin{figure}[b!]
    \centering
        \includegraphics[width=8.5cm]{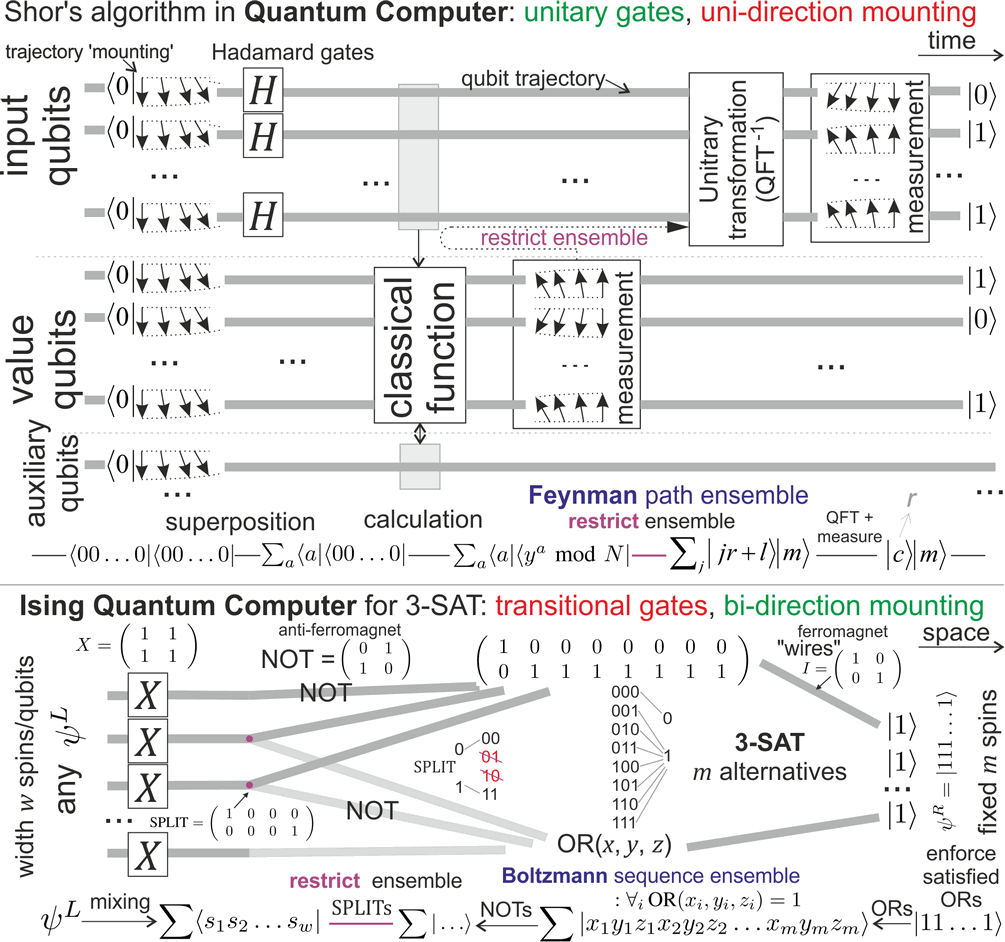}
        \caption{\textbf{Top}: quantum subroutine for Shor's factorization algorithm from \cite{fqm}. We start with initial state prepared as all zeros, then Hadamard gates $H$ produce ensemble of all possible inputs (exponential number), on which there is calculated classical function. Measurement of output of this function restricts the original ensemble to only inputs giving the measured output. Mathematically this restriction is to a periodic set of inputs, its Fourier transform (QFT) allows to obtain this period, which is helpful for finding the factors. \textbf{Bottom}: Ising Quantum Computer (IQC) schematic application to 3-SAT problem (NP-complete). Instead of unitary gates, we have transitional gates with nonnegative coefficients here. However, while QC allows to fix only initial state in the past - only getting a random measurement outcome in the future, spatial realization of IQC allows to choose state/amplitude from both directions: fix left to $\psi^L$ and right to $\psi^R$. This way in the presented 3-SAT example scheme, applying mixing $X$ to all spins/qubits can transform any $\psi^L$ into Boltzmann superposition of all $2^w$ inputs. Connecting them to all 3-SAT alternatives, and enforce all their outcomes to 1, through Boltzmann (uniform here) sequence ensemble should enforce initial mixing to values satisfying all alternatives. Spatial "wires" for such spins require ferromagnetic interaction enforcing identical values in neighboring spins, gates require specific interactions between e.g. 3 spins in previous layer and 1 in the following for $\textrm{OR}(x,y,z)$. }
        \label{IQC}
\end{figure}

That paper also discusses such Feynman path ensemble view on quantum algorithms - especially Shor's factorization, visualized here in Fig. \ref{IQC}. While in quantum computers we rather(?) can only ask about the final states by measuring them, Ising model is generally assumed to allow for realization of Boltzmann path ensembles - what seems a bit weaker computationally, but compensates it with very powerful additional possibility of being able to mount these trajectories in both directions: left and right (instead of only in the past in quantum computers) - suggesting to consider (Wick-rotated) Ising Quantum Computers, briefly introduced here, to be studied in more details in following articles.

The basic idea is using let say width $w$ length $l$ lattice of spins, having a way to enforce boundary situations (in 'length' direction): amplitude from left $\psi^L$ and right $\psi^R$ (real nonnegative, sums of squares are 1) e.g. through weakening interactions on left and acting with strong magnet on right in the discussed example. In contrast to QC having complex amplitudes, this time they are real nonnegative and  normalized to 1 sum of squares, there is no interference.

Inside such e.g. size $w\times l$ rectangular lattice of spins, we would like to encode an instance of a problem to be solved, e.g. through printed layers and external fields. Then assuming that physics indeed uses Boltzmann distribution among possible sequences, chosen configuration of spins should contain a solution (or a hint) - somehow reading it should help with solving given instance of the problem.

So assume we can control transition matrices inside this $w\times l$ rectangle, this time such matrix can very between $l$ layers of width $w$ stripes: we can choose $\{M^x\}_{x=1..l-1}$ sequence of transition matrices e.g. by printed layers and external field to encode instance of a problem. While for constant transition matrix we had worked with dominant eigenvector, it becomes more complex for varying, analogously to time-dependent quantum mechanics, discussed from MERW perspective in Section 5 of \cite{MERW2}. Such Boltzmann path ensemble satisfies for $\gamma=(\gamma^1\ldots \gamma^l)$ sequence of width $w$ stripes:
\be \textrm{Pr}(\gamma)\propto \psi^L_{\gamma^1}\, M^1_{\gamma^1\gamma^2 }\,M^2_{\gamma^2 \gamma^3} \ldots M^{l-1}_{\gamma^{l-1} \gamma^{l}} \, \psi^R_{\gamma^l} \ee
what requires normalization: dividing by sum over all such sequences, for constant transition matrix becoming (\ref{seq1}).

While in quantum computers these intermediate matrices are unitary: with spectrum in unitary circle, here they are "transitional": theoretically with any with real nonnegative coefficients - corresponding to $\exp(-\beta E)$ where $E$ is energy of a given edge in assumed Boltzmann sequence ensemble. Using 0/1 transition matrix corresponds to uniform sequence ensemble (original MERW, forbidden edges have $E=\infty$).

As in quantum computers (QC), this sequence of spins/qubits: hiding $2^w$ possibilities, can be imagined as a tensor product. For systematic design we can build the transition matrices from gates for one or a few spins/qubits: this time transitional gates with nonnegative coefficients instead of QC unitary gates. For example quantum computers use one-qubit Hadamard gate $H$ to prepare entanglement, here it can be replaced with mixing gate $X$:
$$H=\frac{1}{\sqrt{2}} \left(\begin{array}{cc}1 & 1 \\ 1 & -1 \\  \end{array} \right)
\qquad\qquad X=\left(\begin{array}{cc}1 & 1 \\ 1 & 1 \\  \end{array} \right) $$
For example construction of violation of Mermin's inequality from \cite{fqm} can be seen that after preparing chosen amplitude in both directions (the same: $\psi_{000}=\psi_{111}=0$, $\psi_{001}=\psi_{010}=\psi_{100}=\psi_{011}=\psi_{110}=\psi_{101}=1/\sqrt{6}$ ), for measurement of let say $A,B$ variables we use mixing matrix $X$ for the unmeasured remaining $C$ variable - summing corresponding amplitudes, then multiply both sums in such Born rule ($\textrm{Pr}(AB)\propto (\psi_{AB0}+\psi_{AB1})^2$), getting $\textrm{Pr}(A=B)=1/5$ leading to violation of the inequality.

Other basic gates that, at least in theory, should be allowed to use in IQC are permutations of values, like NOT for a given qubit. There are also allowed controlled gates, e.g. control-NOT, control-X, etc.

Spatial realization in Ising-like system allows for additional advantage of bi-directional mounting: fixing amplitudes from both directions, what might allow to directly attack even NP-complete problems like 3-SAT, as sketched in Fig. \ref{IQC}. (However, using SPLIT $=\left(1\ 0\ 0\ 1\right) $ instead seems sufficient, allowing also for uni-directional realization)

In 3-SAT (satisfiability), an instance of a problem is given by size $m$ set of alternatives of triples of some of used variables, some of which can be negated (NOT). The question is if values of variables can be chosen to satisfy all $m$ alternatives (triples). So imagine our width $w$ is the number of such variables, we can prepare $\psi^L$ as (Boltzmann) entanglement of all $2^w$ possibilities, e.g. by starting with any $\psi^L$ and processing all spins/qubits with $X$ gates (or take Ising $J=0,\mu=0$). Now we connect corresponding triples of variables (some NOT-ed) to $m$ OR gates - returning 0 only for '000':
$$\textrm{OR}(x,y,z)^T= \left(\begin{array}{cccccccc}1 & 0 &0 &0 &0 &0 &0 &0  \\ 0 & 1& 1& 1& 1& 1& 1& 1 \\  \end{array} \right) $$
further using $\psi^R$ of size $m$ with fixed all values to 1 (e.g. strong magnetic field), Boltzmann (uniform here) distribution among paths should restrict initial ensemble to values satisfying all alternatives. We can assume that there is exactly one of them (still NP-complete problem), in which case somehow reading these spins we would get solution to our problem.

Obviously, practical realization of such setting has many technical challenges, and Ising assumption of Boltzmann distribution among possible sequences might be only an approximation. However, standard QC also have unimaginably difficult challenges which for practicality might never be overcame, and are essentially different from of discussed IQC - providing looking promising alternative direction.

Notice that discussed Wick-rotated quantum computing is essentially different from adiabatic quantum computing - searching global minimum of Hamiltonian, which for hard problems has exponential number of local minima (otherwise we could quickly search them in classical computer). In contrast, as Shor algorithm, the discussed one directly exploits path ensembles.

\end{document}